\documentclass[noshowpacs,pre,preprintnumbers,twocolumn,superscriptaddress,amsmath,amssymb,floatfix]{revtex4}
\usepackage[caption=false]{subfig}
\usepackage{graphicx}
\usepackage{color}
\usepackage{placeins}
\usepackage{tikz}
\usepackage{mathtools}
\usepackage{amsthm}
\usepackage{romannum}
\usepackage{color}
\usepackage{xkeyval,xcolor}
\usepackage{amsmath}
\usepackage{commath}
\usepackage{hyperref}

\newcommand{\er}{Erd\H{o}s-R\'{e}nyi }
\newcommand{\kk}{\langle k \rangle}
\newcommand{\kkk}{\langle k^2 \rangle}
\newcommand{\x}{{\mathbf {x}}}
\newcommand{\av}[1]{\langle #1 \rangle}
\newcommand{\blue}[1]{{\color{blue} #1}}

\newcommand{\F}{\mathcal{F}}
\newcommand{\s}{\mathcal{S}}
\newcommand{\D}{\mathcal{D}}

\begin{document}

\title{\Large\color{blue} External field and critical exponents \\[5pt] in controlling dynamics on complex networks}
\author{Hillel Sanhedrai}
\affiliation{Department of Physics, Bar-Ilan University, Ramat Gan, Israel}
\author{Shlomo Havlin}
\affiliation{Department of Physics, Bar-Ilan University, Ramat Gan, Israel}
\date{\today}

\begin{abstract} 
	Dynamical processes on complex networks, ranging from biological, technological and social systems, show phase transitions between distinct global states of the system.
	Often, such transitions rely upon the interplay between the structure and dynamics that takes place on it, such that weak connectivity, either sparse network or frail interactions, might lead to global activity collapse, while strong connectivity leads to high activity.
	Here, we show that controlling dynamics of a fraction of the nodes in such systems acts as an external field in a continuous phase transition. As such, it defines corresponding critical exponents,
	both at equilibrium and of the transient time. We find the critical exponents for a general class of dynamics using the leading orders of the dynamic functions. By applying this framework to three examples, we reveal distinct universality classes.	
\end{abstract}

\maketitle

\pagenumbering{arabic}

\blue{\section{Introduction}}

Phase transitions (PT) have attracted tremendously broad research ranging from states of matter through superconductivity to ferromagnetism and many other systems  \cite{Stanley1971,domb2000phase,yeomans1992statistical}.
Particularly, a special focus has been given to the behavior near the transition, revealing various critical phenomena including critical exponents and universality classes \cite{Stanley1971,domb2000phase,yeomans1992statistical}. 

In the field of complex networks, a percolation PT has been widely explored where the order parameter is the relative size of the giant component while the tuning parameter is the occupation probability \cite{cohen2000prl,cohen2001breakdown,Dorogovtsev2008,Buldyrev2010,Stauffer1992}.

Here we consider the PT of dynamical complex systems caused by structural variation \cite{Gao2016,strogatz2018nonlinear,katok1997introduction,perrings1998resilience,May1977}, where a too weakly connected network, even if still connected, cannot function normally as if it is strongly connected. Therefore, in this context, the activity state of the system is regarded as the order parameter, and the tuning parameter is the connectivity as we define below.

The impact of controlling nodes' activity on network dynamics has been investigated from several aspects, ranging from controllability theory \cite{liu2011controllability,liu2016control,whalen2015observability}, through propagation patterns of small perturbations across the network \cite{Barzel2013a,Harush2017,Hens2019}, to global effects on the system state \cite{Cornelius2013,sanhedrai2022reviving,sanhedrai2022sustaining}. 

In this paper, we focus on the critical behavior, and we show that the external intervention in dynamics of a fraction of nodes can be regarded as an external field in continuous PT, analogously to an external magnetic field in a ferromagnetic PT \cite{hook2013solid}. 

We define new critical exponents besides those which are already known \cite{grassberger1983critical,biswas2012disorder}, and we find the values of the exponents for a \emph{general form of dynamics}. By applying our general framework to several dynamical models we reveal distinct universality classes which indicate essential distinct responses to control around criticality. 
The known critical exponents of ferromagnetism PT for mean-field are obtained as a particular case of our general results.

In addition to the critical exponents of the equilibrium, we define and analyze also critical exponents related to the transient time towards equilibrium \cite{yahata1969critical,wang1995study}. Different from equilibrium critical exponents which can be formed by only two independent exponents, the three transient exponents are independent. Nevertheless, all exponents can be derived from three independent exponents.

\vspace{1cm}

\begin{figure*}[ht]
	\centering
	\includegraphics[width=0.99\linewidth]{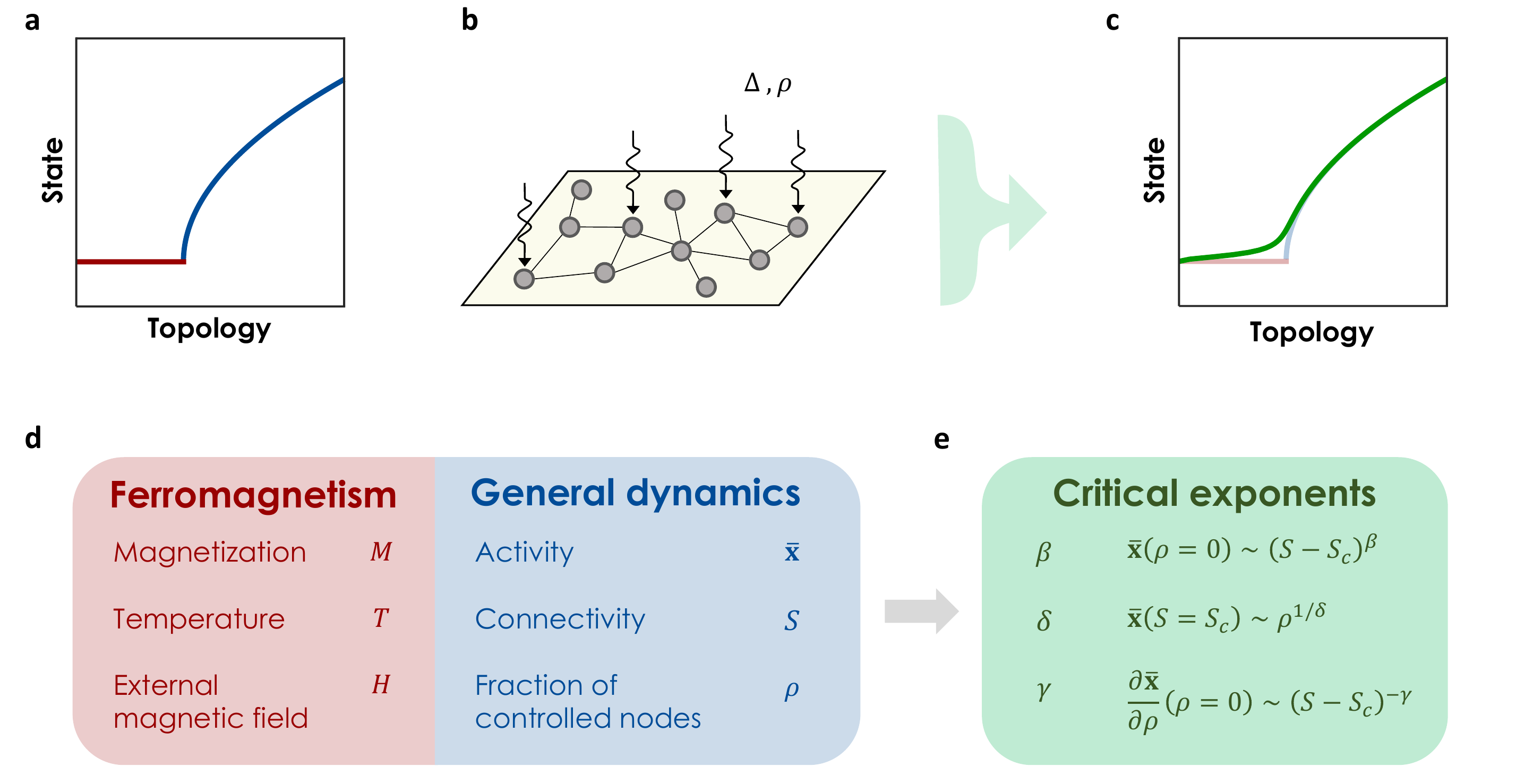}
	\caption{{\bf Control as an external field.} 
		(a) Dynamical systems as epidemic spread, gene regulation and opinion dynamics often show a continuous transition between an inactive state (red) for low connectivity, and an active state (blue) for high connectivity. We consider such systems in this paper.
		(b) We select randomly a fraction $\rho$ of nodes, and force them to have a constant activity $\Delta>0$.
		(c) The effect of the control described in (b) is that the activity of the system increases and the critical point is eliminated. This effect is analogous to that of an external magnetic field in a ferromagnetic phase transition.
		(d) We define for general dynamical systems, analogously to ferromagnetism, an order parameter (activity as magnetization), tuning parameter (connectivity as temperature) and external field (a fraction of nodes as an external magnetic field).
		(e) The definitions in (d) yield the corresponding definitions for critical exponents.
	  }
	\label{fig:illustration}
\end{figure*}

\blue{\section{Controlling system dynamics}}

To analyze the impact of an intervention in network dynamics we rely upon a general framework \cite{Barzel2013a,Harush2017,Hens2019} to model nonlinear dynamics on networks.
Consider a system consisting of $N$ components (nodes) whose activities $x_i$ ($i=1,2,...,N$) follow the Barzel-Barab\'asi equation \cite{Barzel2013a},
\begin{equation}
	\dod{x_i}{t} = M_0(x_i)+ \omega\sum\limits_{j=1}^{N}A_{ij}M_1(x_i)M_2(x_j).
	\label{Dynamics}
\end{equation}
The first function, $M_0(x_i)$ captures node $i$'s self-dynamics, describing mechanisms such as protein degradation \cite{Barzel2011} (cellular), individual recovery \cite{Dodds2005,PastorSatorras2015} (epidemic) or birth/death processes \cite{Gardner2000} (population dynamics). The product $M_1(x_i)M_2(x_j)$ describes the $i,j$ interaction mechanism, \textit{e.g.}, genetic activation \cite{Alon2006,Karlebach2008,schreier2017exploratory}, infection \cite{Dodds2005,PastorSatorras2015} or symbiosis \cite{Holling1959}. The connectivity matrix $A$ captures the interactions (links) between the nodes, \textit{i.e.}\ the network. An element $A_{ij}$ equals $1$ if there is a link (interaction) between nodes $i$ and $j$ and $0$ otherwise. We consider here a matrix $A$ which is symmetric and obeys the configuration model framework, that is a random network with a given degree distribution $p_k$. The strength of the interactions is governed by the positive uniform parameter $\omega$.

As an external intervention in network dynamics we consider the following simple control. We \emph{force} a set of nodes, $\F$ (a fraction $\rho$ of the system), to have a constant high activity value $\Delta$ (Fig.\ \ref{fig:illustration}b), while all the rest in the complementary set, $\D$, are governed by the original dynamics. 
Thus, such a forced system obeys the set of equations,
\begin{equation}
	\left\{
	\begin{array}{cclr}
		x_i &=& \Delta & i \in \mathcal{F} ,
		\\[7pt]
		\dod{x_i}{t} &=& M_0(x_i) + \omega \displaystyle \sum_{j = 1}^N A_{ij} M_1(x_i) M_2(x_j) \hspace{5mm} & i \in \D .
	\end{array}
	\right. 
	\label{ForcedDynamics}
\end{equation} 

\vspace{4mm} 
\noindent
In this study, we assume that the set of controlled nodes, $\F$, is selected randomly.
Next, we aim to track the states of the unforced nodes, \textit{i.e.}\ the set $\D$.  

Using a mean field approximation \cite{Gao2016,sanhedrai2022sustaining} (see SI Section 3), and considering a \emph{random} selection of controlled nodes, we obtain for the steady states \cite{sanhedrai2022sustaining},
\begin{equation}
	\s = \frac{ -M_0(\bar{\x}) }{(1-\rho) M_1(\bar{\x})M_2(\bar{\x}) + \rho M_1(\bar{\x}) M_2(\Delta)},
	\label{sX}
\end{equation}
where the order parameter, $\bar{\x}$, indicating the system state, is the average activity over all the neighbors within the dynamic set, $\D$, defined by
\begin{equation} \label{eq:xBar}
	\bar{\x} = \frac{1}{|\D|\av{k_{\D\to\D}}} \sum_{j \in \D} k_j^{\D\to\D}  x_j .
\end{equation}
The quantity $k_j^{\D\to\D}$ denotes the number of free neighbors, \textit{i.e.}\ within $\D$, of a free node $j\in\D$ .
The connectivity, $\s$, in Eq.\ \eqref{sX}, is defined as
\begin{equation}
	\s = \omega \kappa,
	\label{S}
\end{equation}
where $\omega$ is the interaction strength from Eq.\ \eqref{Dynamics}, and $\kappa$ is the average neighbor degree over the whole network, $\kappa = \kkk/\kk$.
Finally, $\rho=|\F|/N$ is the fraction of controlled nodes, and $\Delta$ is the value of forcing, Eq.\ \eqref{ForcedDynamics}. 

Eq.\ \eqref{sX} provides a relation between the system state, $\bar{\x}$, and the connectivity $\s$. By substituting $\rho=0$ we get the phase diagram of the free system. In Fig.\ \ref{fig:illustration}a we show a typical result of Eq.\ \eqref{sX} with $\rho=0$ which we discuss in this paper. The obtained curve exhibits a continuous phase transition (PT) between an inactive (red, $\bar{\x}=0$) and an active (blue, $\bar{\x}>0$) states for a \emph{free} system without any external control.

\hspace{1cm}

\blue{\section{External field analogy}}

Forcing a system as described in Eq.\ \eqref{ForcedDynamics} and illustrated in Fig.\ \ref{fig:illustration}b, we obtain the typical phase diagram presented in Fig.\ \ref{fig:illustration}c which is constructed by Eq.\ \eqref{sX} with $\rho>0$.
One can see that controlling the system acts as an external field in a continuous PT \cite{Stanley1971,Stauffer1992,bunde2012fractals}. It makes the curve smooth and eliminates the phase transition (green curve). Therefore, the system stable state, $\bar{\x}$, the connectivity, $\s$, and the fraction of controlled nodes, $\rho$, in PT of dynamical systems, are being an analogy of magnetization, temperature and external magnetic field respectively in ferromagnetic PT, Fig.\ \ref{fig:illustration}d.  
Hence, we define the fraction of forced nodes, $\rho$, as the strength of the external field in our problem. 
Correspondingly, we define the following critical exponents,
\begin{eqnarray}
	\label{betaDef}
	\bar{\x}(\s\to\s_c^+,\rho=0) & \sim & {(\s-\s_c)}^{\beta} ,
	\\[8pt]
	\label{deltaDef}
	\bar{\x}(\s=\s_c,\rho\to0) & \sim & {\rho}^{1/\delta} ,
	\\[8pt]
	\label{gammaDef}
	\chi(\s\to\s_c^+,\rho=0) & \sim & {(\s-\s_c)}^{-\gamma} ,
\end{eqnarray} 
where the susceptibility, $\chi$, is defined by
\begin{equation}
	\chi=\dpd{\bar{\x}}{\rho} \bigg|_{\rho=0} .
	\label{chiDef}
\end{equation} 
In the next Section, we show that these scaling relations are satisfied, and we find the critical exponents generally for given dynamics captured by the functions $M_{0,1,2}$ in Eq.\ \eqref{Dynamics}. 

\vspace{1cm}

\begin{figure*}[ht]
	\centering
	\includegraphics[width=0.99\linewidth]{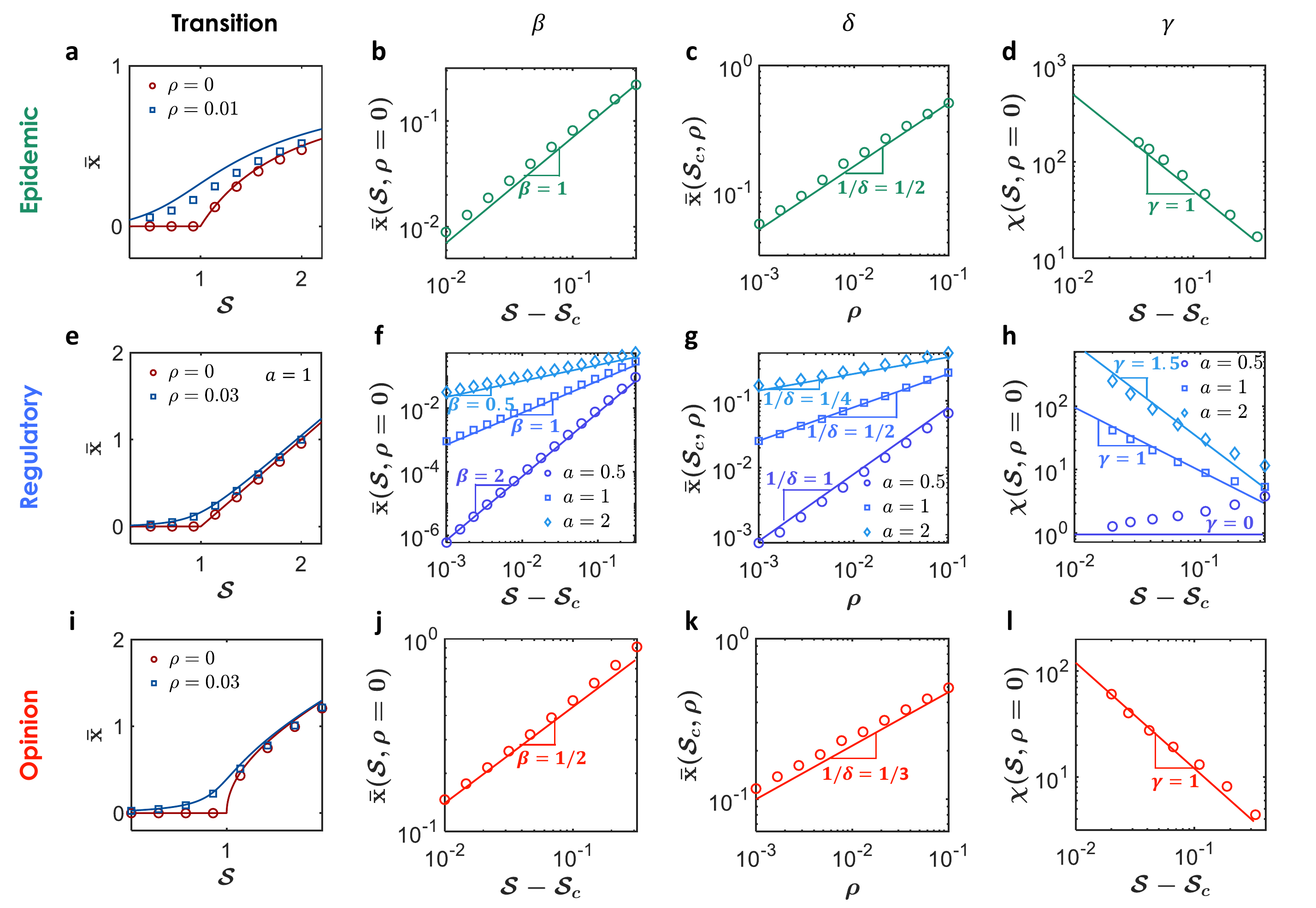}
	\caption{{\bf Equilibrium critical exponents.} 
		We apply our general results in Eqs.\ \eqref{beta}-\eqref{gamma} to three dynamical processes, epidemic, regulatory and opinion dynamics, which exhibit different critical exponents as shown in Table \ref{table}. All simulation (symbols) were performed on \er networks of size $N=10^4$ with average neighbor degree $\kappa=10$, and varying weight $\omega$ and support our theoretical predictions (lines).
		(a)-(d) Epidemic dynamics, Eq.\ \eqref{sis}.
		(a) The phase diagram obtained from Eq.\ \eqref{sX} (lines) with $\rho=0$ for a free system and $\rho=0.01$ for a forced system compared to simulation results. This system shows a continuous phase transition that gets removed by control analogous to an external field effect.
		(b)-(d) The predicted critical exponents derived from Eqs.\ \eqref{beta}-\eqref{gamma} as detailed in Sec.\ \ref{sec:examples} and given in Table \ref{table} are supported by simulations.
		(e)-(h) Regulatory dynamics, Eq.\ \eqref{MM}.
		(e) The phase diagrams for free and forced systems show the same effect as of an external field in continuous PT. Here we set $a=1$ in Eq.\ \eqref{MM}.
		(f)-(h) The predicted critical exponents derived from Eqs.\ \eqref{beta}-\eqref{gamma} as detailed in Sec.\ \ref{sec:examples} and summarized in Table \ref{table} are supported by simulations.
		(i)-(l) Opinion dynamics, Eq.\ \eqref{opinions}.
		(i) The same as (e) for opinion dynamics.
		(j)-(l) The critical exponents are different from other examples and identical to the well-known mean-field exponents of Ising model since a very similar model to opinion dynamics is assumed to describe spin dynamics.
	}
	\label{fig:equilibrium}
\end{figure*}

\blue{ \section{General derivation of the critical exponents} }

We are interested in finding the critical exponents for the general dynamic functions $M_{0,1,2}$ appearing in Eq.\ \eqref{Dynamics}. 
As aforesaid, we consider dynamics which have a stable state $\bar{\x}=0$ below criticality.
We also consider a continuous PT, such that the active ($\bar{\x}>0$) state approaches zero when $\s\to\s_c^+$, thus we analyze Eq.\ \eqref{sX} in the limit of $\bar{\x}\to0$. To this end, we assume that the dynamical functions have Hahn expansions as power series \cite{Hahn1995},
\begin{equation}
	\begin{aligned}
		M_0(x) = \sum\limits_{n=0}^{\infty}a_nx^{\Gamma _n} ,
		\\
		M_1(x)M_2(x) = \sum\limits_{n=0}^{\infty}b_nx^{\Pi _n} ,
		\\
		M_1(x) = \sum\limits_{n=0}^{\infty}c_nx^{\Lambda _n},
	\end{aligned}
	\label{expansions}
\end{equation}
where $n$ is an index running over all powers that appear in the expansion either the series is infinite or finite. The exponents $\Gamma _n$, $\Pi _n$ and $\Lambda_n$ are increasing non-negative series, and differently from the Taylor series they can be fractional.

The above functions should satisfy the characters of our problem as mentioned above, that is a continuous transition from an inactive stable state, $\bar{\x}=0$, for a weak connectivity, $\s<\s_c$, to an active stable state, $\bar{\x}>0$, for a strong connectivity, $\s>\s_c$. In addition, a controlled system should act as if under an external field. To this end (see derivation in SI Section 4), the lead exponents should fulfill  
$\Gamma_0 =\Pi_0 >\Lambda_0$,
and the lead coefficients should satisfy
$a_0<0$, $b_0>0$, and
$c_0M_2(\Delta)>0$. For the non-leading terms see SI Section 5.1.

As we expand Eq.\ \eqref{sX} at the limit of $\s\to\s_c^+$, $\rho\to0$, and thus also assume $\bar{\x}\to0$, by using the expansions in Eq.\ \eqref{expansions}, we obtain that the scaling relations of Eqs.\ \eqref{betaDef}-\eqref{gammaDef} hold, and the critical exponents depend on the lead exponents of the dynamical functions as (see SI Section 5)
\begin{eqnarray}	
	\beta & = \dfrac{1}{m-\Gamma_0} ,
	\label{beta}
	\\[8pt]
	\delta & = m-\Lambda_0 ,
	\label{delta}
	\\[8pt]
	\gamma & = \dfrac{m-\Lambda_0-1}{m-\Gamma_0} ,
	\label{gamma}
\end{eqnarray}
where 
\begin{equation}
	m = \min\{\Gamma_1,\Pi_1\}.
\end{equation}
If $\Gamma_1$ does not exist, then simply $m=\Pi_1$, and vice versa.
One can see that these three critical exponents obey the known scaling relation, the Widom's identity, $\beta(\delta-1)=\gamma$, implying that there are only two independent exponents which form the third one as already well-known.

In Fig.\ \ref{fig:equilibrium} we show simulation results that agree with our theoretical predictions for the critical exponents for epidemic, regulatory and opinion dynamics, for \er random network. In Sec.\ \ref{sec:examples} we analyze specifically each dynamics and find its critical exponents summarized in Table \ref{table}.
As one can see, different dynamics exhibit different critical exponents which indicate they belong to distinct universality classes. The regulatory dynamics exponents depend on a parameter $a$. Setting $a$ to 1 will equalize its exponents to those of epidemic dynamics. In contrast, there is no value of $a$ that would equalize the exponents to those of opinion dynamics.

\begin{figure*}[ht]
	\centering
	\includegraphics[width=0.8\linewidth]{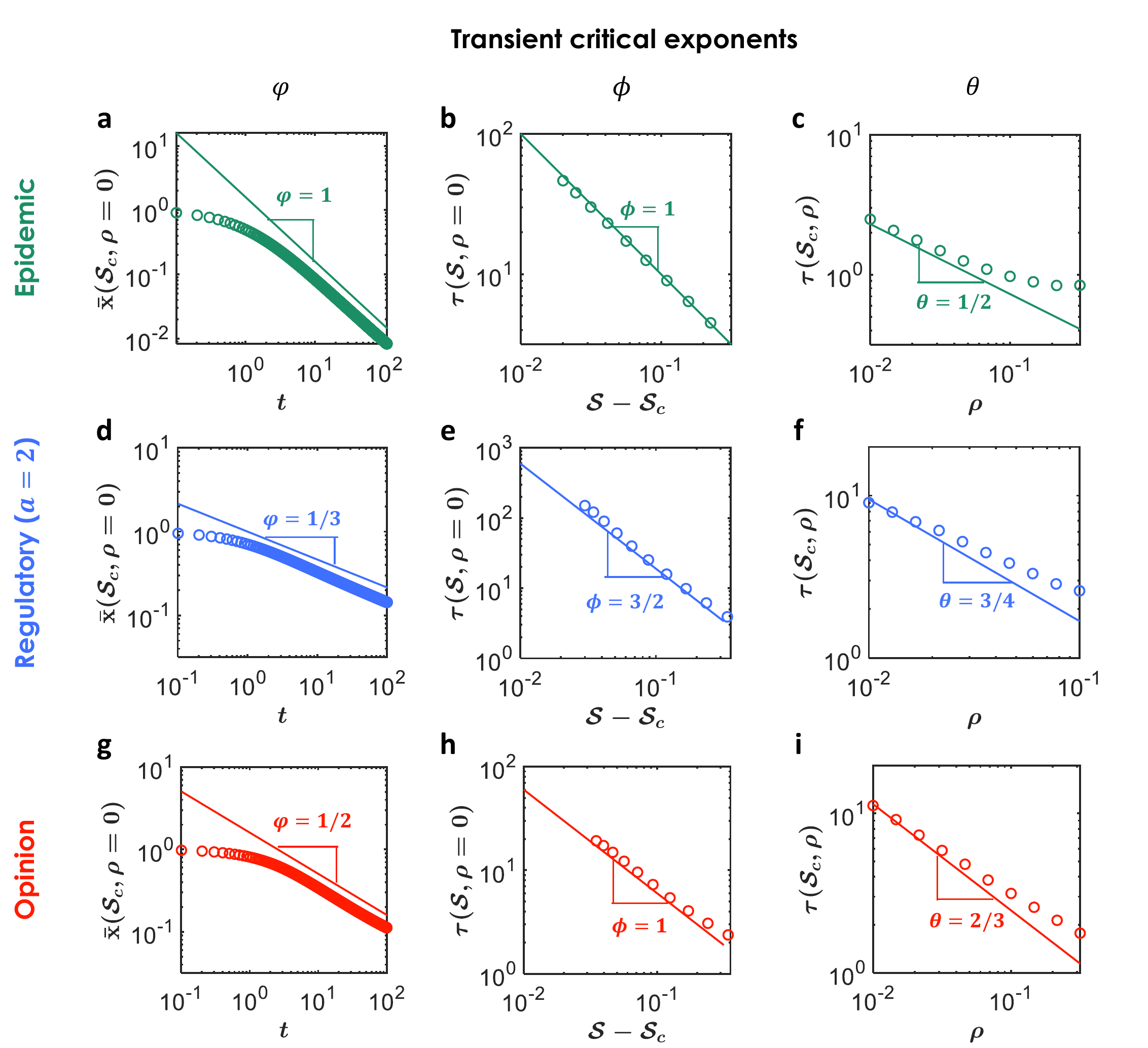}
	\caption{{\bf Transient critical exponents.} 
		The predicted transient critical exponents derived for general dynamics from Eqs.\ \eqref{varphi}-\eqref{theta} are applied to three dynamical examples as in Sec.\ \ref{sec:examples}, and supported by computer simulations (symbols). Lines represent theoretical results.
		(a)-(c) Results for epidemic dynamics, Eq.\ \eqref{sis}.
		(d)-(f) Regulatory dynamics captured by Eq.\ \eqref{MM} with $a=2$.
		(g)-(i) Opinion dynamics, Eq.\ \eqref{opinions}. All simulation were performed on \er networks of size $N=10^4$ with average neighbor degree $\kappa=10$, and varying weight $\omega$.
	}
	\label{fig:transient}
\end{figure*}

\vspace{1cm}

\blue{\section{Transient critical exponents}}

After we analyzed the system state at equilibrium, we further analyze the time it takes for the system to relax towards equilibrium.
To this end, we go one step back before Eq.\ \eqref{sX}, (see SI Section 3) to use the mean-field \emph{dynamic} equation, which follows the evolution in time of the average state of the system, \cite{Gao2016,sanhedrai2022sustaining}
\begin{equation}
	\dod{\bar{\x}}{t}=  M_0(\bar{\x}) + \s M_1(\bar{\x})\left((1-\rho) M_2(\bar{\x}) + \rho  M_2(\Delta)\right).
	\label{dXdt}
\end{equation}
Similarly to the above analysis of Eq.\ \eqref{sX} for finding the critical exponents of the steady state, here we analyze Eq.\ \eqref{dXdt} in the limit of large $t$, $\s\to\s_c^+$ and $\rho\to0$, and thus also $\bar{\x}\to0$, to obtain the critical exponents of the transient towards the stable state (see SI Sec.\ 6 for details).
Correspondingly, we define additional three exponents related to relaxation time. At criticality, \textit{i.e.}\ at $\s=\s_c$, $\rho=0$, we find a power law convergence with the exponent $\varphi$ defined by, 
\begin{equation} \label{varphiDef}
	\bar{\x}(\s=\s_c,\rho=0,t\to\infty) \sim {t}^{-\varphi} .
\end{equation}
This power-law convergence is valid for $m>1$ (see SI Sec.\ 6) which is the case in all our examples.
When $m=1$, the decay at criticality is exponential, and if $m<1$, the system relaxes in a finite time.

Above criticality, that is for $\s>\s_c$ or $\rho>0$, there is an exponential decay, with a decay time $\tau$, which depends on both $\s$ and $\rho$,
\begin{equation}
	\bar{\x}(\s,\rho,t\to\infty) \sim \exp(-t/\tau(\s,\rho)).
\end{equation}
However, when we approach the critical point, we get that the typical relaxation time $\tau$ diverges because the decay becomes as power law rather than exponential. We track two paths in $(\s,\rho)$-space towards the critical point $(\s_c,0)$, vertical and horizontal, and define respectively, 
\begin{eqnarray}
	\label{phiDef}	
	\tau(\s\to\s_c^+,\rho=0) & \sim & {(\s-\s_c)}^{-\phi},
	\\[8pt]
	\label{thetaDef}
	\tau(\s=\s_c,\rho\to0) & \sim & {\rho}^{-\theta}.
\end{eqnarray}
By expanding Eq.\ \eqref{dXdt} close to criticality (see SI Sec.\ 6), we find the transient critical exponents via the expansions of the dynamic functions, Eq.\ \eqref{expansions},
\begin{eqnarray}
	\label{varphi}
	\varphi & = \dfrac{1}{m-1} ,
	\\[8pt]
	\label{phi}
	\phi & = \dfrac{m-1}{m - \Gamma_0}	,
	\\[8pt]
	\label{theta}
	\theta & = \dfrac{m-1}{m - \Lambda_0} .
\end{eqnarray}
Note that these three transient exponents are independent in contrast to the above equilibrium exponents. However, all six exponents can be obtained from three independent exponents, since they all depend on only three leading terms of the dynamical functions, $m$, $\Gamma_0$ and $\Lambda_0$.

In Fig.\ \ref{fig:transient} we show the results of simulations on \er network for our three dynamical examples, which exhibit good agreement with our theoretical predictions.

\vspace{1cm}

\blue{\section{Applications}
\label{sec:examples}}

In this Section, we apply our general analysis to three examples of dynamics (see also SI Sec.\ 7), which fulfill our demand for a continuous PT, that behave under external control as a ferromagnetic matter under an external magnetic field as discussed before.

\vspace{7mm}

\blue{\subsubsection{Epidemic}}

As our first example, we consider the Susceptible-Infectious-Susceptible (SIS) model for epidemic spreading \cite{Barthelemy2005,Hufnagel2004,Dodds2005}. In this model, Eq.\ \eqref{Dynamics} takes the form,
\begin{equation}
	\dod{x_i}{t} = -\alpha x_i + \omega\sum\limits_{j=1}^{N}A_{ij}(1-x_i)x_j,
	\label{sis}
\end{equation}
where $x_i(t)$ represents the probability of agent $i$ to be infectious. The first term on the rhs indicates the probability of recovering with recovery rate $\alpha$. We set, without loss of generality, the recovery rate to be $\alpha=1$. The second term represents the likelihood for node $i$ to get infected by its neighbors with rate $\omega$. This model is mapped to our form in Eq.\ \eqref{Dynamics} by
$M_0(x_i)=-x_i$, $M_1(x_i)=1-x_i$, and $M_2(x_j)=x_j$. Therefore, $\Gamma_0=1$ and $\Gamma_1$ does not exist, $\Pi_0=1$ and $\Pi_1=2$, $\Lambda_0=0$, and $m=2$. Hence, the equilibrium critical exponents, using Eqs.\ \eqref{beta}-\eqref{gamma}, are $\beta=1$, $\delta=2$, and $\gamma=1$. The transient critical exponents, obtained by Eqs.\ \eqref{varphi}-\eqref{theta}, are $\varphi=1$, $\phi=1$ and $\theta=1/2$.
As shown in Fig.\ \ref{fig:equilibrium}a-d and Fig.\ \ref{fig:transient}a-c, the predicted exponents are supported by computer simulations on \er networks with neighbor degree $\kappa=10$. 

\vspace{7mm}

\blue{\subsubsection{Regulatory}}

Our second example is gene regulatory dynamics governed, according to Michaelis-Menten (MM) model \cite{Karlebach2008}, by
\begin{equation}
	\dod{x_i}{t} = -Bx_i^a + \omega\sum\limits_{j=1}^{N}A_{ij}\frac{x_j^h}{1+x_j^h}.
	\label{MM}
\end{equation}
Under this framework, $M_0(x_i) = -B x_i^a$, describing degradation ($a = 1$), dimerization ($a = 2$) or a more complex bio-chemical depletion process (fractional $a$), occurring at a rate $B$; without loss of generality we set here $B = 1$. The activation interaction is captured by the Hill function of the form $M_1(x_i) = 1$, $M_2(x_j) = x_j^h/(1+x_j^h)$, a \textit{switch-like} function that saturates to $M_2(x_j) \rightarrow 1$ for large $x_j$, representing node $j$'s positive, albeit bounded, contribution to node $i$ activity, $x_i(t)$. 

The appropriate case for a continuous transition is when $a=h$ to satisfy the relation mentioned above, $\Gamma_0=\Pi_0$. Thus, $\Gamma_0=a$ and $\Gamma_1$ does not exist, $\Pi_0=a$ and $\Pi_1=2a$, $\Lambda_0=0$ and $m=2a$. Hence, the equilibrium critical exponents, using Eqs.\ \eqref{beta}-\eqref{gamma}, are $\beta=1/a$, $\delta=2a$ and $\gamma=2-1/a$. The transient critical exponents, obtained by Eqs.\ \eqref{varphi}-\eqref{theta}, are $\varphi=1/(2a-1)$, $\phi=2-1/a$ and $\theta=1-1/(2a)$. The simulation results compared to these predictions are presented in Fig.\ \ref{fig:equilibrium}e-h and Fig.\ \ref{fig:transient}d-f.
An interesting result is for $a=1/2$ or less when some of the exponents become zero or even negative. See Fig.\ \ref{fig:equilibrium}h for $a=1/2$ (circles) where, as predicted, the susceptibility does not diverge at criticality in contrast to all other examples. This indicates an essential change in the effect of the control. While for other cases the response at criticality to control is very high, in regulatory dynamics with $a=1/2$, the system responds uniformly to control.

\renewcommand{\arraystretch}{2.5}

\begin{table*}[ht]
	\centering
	\begin{tabular}{l l c c c c c c} 
		\hline \hline
		\bf Dynamics  \hspace{10mm} & \bf Model \hspace{10mm} & \hspace{5mm} \boldmath $\beta$ \hspace{5mm} & \hspace{5mm} \boldmath $\delta$ \hspace{5mm} & \hspace{5mm}  \boldmath $\gamma$ \hspace{5mm} & \hspace{5mm} \hspace{5mm} \boldmath $\varphi$ \hspace{5mm} & \hspace{5mm} \boldmath  $\phi$ \hspace{5mm} &  \hspace{5mm} \boldmath $\theta$ \hspace{5mm} \\[4pt] 
		\hline \hline		
		Epidemic & SIS & 1 & 2 & 1& \hspace{5mm} 1 & 1 & $\dfrac{1}{2}$ \\[4pt] 
		\hline
		Regulatory & MM & $\dfrac{1}{a}$ & $2a$ & $2-\dfrac{1}{a}$ & \hspace{5mm} $\dfrac{1}{2a-1}$ & $2-\dfrac{1}{a}$ & $1-\dfrac{1}{2a}$ \\ [4pt]
		\hline
		Opinion & BLSS & $\dfrac{1}{2}$ & 3 & 1 & \hspace{5mm} $\dfrac{1}{2}$ & 1 & $\dfrac{2}{3}$ \\[4pt]
		\hline
	\end{tabular}
	\caption{All critical exponents found in this paper for three examples of dynamics, including the exponents related to the steady state ($\beta$, $\delta$ and $\gamma$) and also the exponents related to the transient towards relaxation ($\varphi$, $\phi$ and $\theta$). The values are derived from Eqs.\ \eqref{beta}-\eqref{gamma} and Eqs.\ \eqref{varphi}-\eqref{theta}.
	}
	\label{table}
\end{table*}

\vspace{7mm}

\blue{\subsubsection{Opinion}}

Our final example is a model (which we call BLSS) for opinion dynamics \cite{baumann2020modeling},
\begin{equation}
	\dod{x_i}{t} = -x_i + \omega\sum\limits_{j=1}^{N}A_{ij} \tanh (\alpha x_j).
	\label{opinions}
\end{equation}
The sign of $x_i$ describes the agent $i$’s qualitative stance towards a binary issue of choice (\textit{e.g.}\ the preference between two candidates). The absolute value of $x_i$ quantifies the strength of this opinion, or the convincing level.
This model treats opinion dynamics as a
purely collective, self-organized process without any intrinsic individual preferences. Hence, the opinions of agents
without social interactions decay toward the neutral state 0, which is ruled by the self-dynamics function, $M_0(x_i)=-x_i$.
The interaction $ij$ is captured by $M_1(x_i)=1$ and $M_2(x_j)= \tanh(\alpha x_j)$. 
This odd nonlinear shape guarantees that an agent $j$ influences others in the direction of its own opinion’s sign,  with a level that increases monotonically with its convincing level,
albeit with saturation since the social influence of extreme opinions is capped.
We set here $\alpha=1$. 
The leading orders of the dynamic functions are, therefore, $\Gamma_0=1$ and $\Gamma_1$ does not exist, $\Pi_0=1$ and $\Pi_1=3$, $\Lambda_0=0$ and $m=3$. Hence, the obtained equilibrium exponents derived from Eqs.\ \eqref{beta}-\eqref{gamma}, are $\beta=1/2$, $\delta=3$ and $\gamma=1$. 
These are the same critical exponents as those of Ising model for ferromagnetism in mean-field \cite{Stanley1971}. This is not surprising since a very similar model is used to model spin dynamics \cite{krapivsky2010kinetic}.
The transient exponents, yielded from Eqs.\ \eqref{varphi}-\eqref{theta}, are $\varphi=1/2$, $\phi=1$ and $\theta=2/3$. These results are supported via computer simulations in Fig.\ \ref{fig:equilibrium}i-l and Fig.\ \ref{fig:transient}g-i.

\vspace{10mm}

\blue{\section{Discussion}}

In this paper, we considered a general class of complex dynamical systems, analyzed by a general framework, to explore the behavior of the system near criticality of a phase transition (PT). This PT captures the interplay between structure and dynamics such that weak connectivity yields a suppressed activity while strong connectivity leads to an active state of the system. Our main focus is on the impact of controlling dynamics of a fraction of nodes on the system state close to criticality. By only leading terms of the dynamic functions, we construct both \emph{equilibrium} critical exponents and \emph{transient} critical exponents. Applying our framework to three examples we reveal distinct universality classes, indicating the essential different effects of external control in these systems.

Yet, several directions are still needed to be explored for going beyond our results. 
Here we assumed that the controlled nodes are selected randomly. However, external control is likely non-random but could be rather targeted. For instance, a localized control, \textit{i.e.}\ a source node and its neighbors and next neighbors and so on, might exhibit a considerably unequal effect. Another reasonable targeted control is by selecting the high-degree nodes. This case becomes more interesting for scale-free (SF) networks which show a diverse degree distribution.
Speaking about SF networks, in our analysis we used a mean-field approach, relying upon an assumption of relatively small fluctuations, which is challenged by SF networks. Thus, networks with a very broad degree distribution demand further analysis.
Finally, here we focused on continuous transitions, however, an extension of our analysis can explore in a similar way abrupt transitions, see Refs.\ \cite{gross2020interconnections,sanhedrai2022sustaining}.

\vspace{10mm}

\noindent
\blue{\bf Acknowledgments}\\
H.S. acknowledges the support of the Presidential Fellowship of Bar-Ilan University,
Israel, and the Mordecai and Monique Katz Graduate Fellowship Program. 
We thank the Israel Science Foundation, the Binational Israel-China Science Foundation (Grant No. 3132/19), the NSF-BSF (Grant No. 2019740), the EU H2020 project RISE (Project No. 821115), the EU H2020 DIT4TRAM, and DTRA (Grant No. HDTRA-1-19-1-0016) for financial support.
\\

\noindent
\blue{\bf Author contributions}\\
Both authors designed the research and wrote the article. H.S.\ performed the analytical derivations and the computer simulations.
\\

\noindent
\blue{\bf Data availability}\\
No datasets were generated or analyzed during the current study.
\\

\noindent
\blue{\bf Code availability}\\
All codes to reproduce, examine and improve our proposed analysis are available at \url{https://github.com/hillel26/ControlDynamicsCriticalExponents.git}.


\def\bibfont{\small}

\bibliographystyle{unsrt}
{ \onecolumngrid
	\linespread{1.4}
	\bibliography{bibliography}

}

\end{document}


\title{\color{blue}  \bf External field and critical exponents in controlling dynamics on complex networks
\\[20pt] 
Supplementary information
\\[30pt]}
\author{Hillel Sanhedrai and Shlomo Havlin}

\maketitle
\tableofcontents
\clearpage
\pagenumbering{arabic}

\section{Dynamics framework}

We consider a class of systems captured by the Barzel-Barab\'asi equation \cite{Barzel2013},

\begin{equation}
\dod{x_i}{t} = M_0(x_i) + \omega \sum_{j = 1}^N \m Aij M_1(x_i) M_2(x_j),
\label{Dynamics}
\end{equation}

\noindent
where $x_i(t)$ is node $i$'s dynamic \textit{activity} ($i = 1,\dots,N$) and the nonlinear functions, $M_0(x)$, $M_1(x)$ and $M_2(x)$, describe the system's intrinsic dynamics, \textit{i.e.} its self-dynamics ($M_0$) and its interaction mechanisms ($M_1, M_2$). The patterns of connectivity between the nodes are captured by the network $\m Aij$, a binary $N \times N$ adjacency matrix, which we take to follow the configuration model framework, namely a random network with an arbitrary degree distribution $p_k$. The strength of all interactions is governed by the weight $\omega>0$, following constructive or attractive  interactions. We do not consider here a competitive interaction or oscillatory coupling functions.   

For instance, in epidemic spreading, according to SIS model \cite{Barthelemy2005,Hufnagel2004,Dodds2005}, where a susceptible node can get infected by an infectious node, and an infectious node might recover and become susceptible again, the probability of each agent $i$ to be infectious, $x_i$, evolves in time as

\begin{equation}
	\dod{x_i}{t} = - \alpha x_i + \omega \sum_{j = 1}^N \m Aij (1-x_i) x_j,
	\label{SIS}
\end{equation}

where $\alpha$ is the recovery rate, and $\omega$ is the infection rate. The product $(1-x_i)x_j$ represents the probability that agent $i$ is susceptible and node $j$ is infectious.
This dynamical model is mapped to the general form of Eq.\ \eqref{Dynamics} by $M_0(x)=-\alpha x$, $M_1(x)=1-x$ and $M_2(x)=x$. The infection rate $\omega$ is the interaction weight in this dynamics.

\clearpage

\section{Free system states}

Before we analyze in the next Section the effect of controlling system dynamics, which is our main goal in this paper, we first introduce the analysis of a free system developed in Ref.\ \cite{Gao2016}.
To follow the global state of the system, we use a mean field approach \cite{Gao2016} to reduce the $N$-dimensional system in Eq.\ \eqref{Dynamics} into a single effective equation. First, we recognize the average over $i$'s neighbors,

\begin{equation}
	\dod{x_i}{t} = M_0(x_i) + \omega M_1(x_i) \bigg( \sum_{j = 1}^N \m Aij \bigg) \frac{\sum_{j = 1}^N \m Aij  M_2(x_j)}{ \sum_{j = 1}^N \m Aij}.
\end{equation}

Next we denote the degree of $i$ as $k_i=\sum_{j=1}^{N}A_{ij}$, and by assuming a mean-field approximation we replace the average over $i$'s neighbors by an average over all neighbors,

\begin{equation}
	\dod{x_i}{t} = M_0(x_i) + \omega M_1(x_i) k_i \frac{\sum_{i,j = 1}^N \m Aij  M_2(x_j)}{ \sum_{i,j = 1}^N \m Aij},
	\label{MF1}
\end{equation} 

which can be written as

\begin{equation}
	\dod{x_i}{t} = M_0(x_i) + \omega M_1(x_i) k_i \frac{\sum_{j = 1}^N k_j  M_2(x_j)}{N\kk }.
	\label{MF2}
\end{equation} 

We denote the weighted average with respect to degree in the interaction term by $\overline{M_2(\x)}$, hence,

\begin{equation}
	\dod{x_i}{t} = M_0(x_i) + \omega M_1(x_i) k_i \overline{M_2(\x)} .
	\label{MF3}
\end{equation} 

By using again a mean-field approach \cite{Gao2016}, we assume that for non-large fluctuations, the leading order satisfies $\overline{M_2(\x)} = M_2(\bar{\x})$, yielding

\begin{equation}
	\dod{x_i}{t} = M_0(x_i) + \omega M_1(x_i) k_i M_2(\bar{\x}) .
	\label{MF4}
\end{equation} 

Finally, we operate an average with respect to degree, as above, on both sides of the last equation, and by using again the mean-field approximation, we obtain

\begin{equation}
	\dod{\bar{\x}}{t} = M_0(\bar{\x}) + \s M_1(\bar{\x})M_2(\bar{\x}),
	\label{MFDynamics}
\end{equation}

where $\bar{\x}$ is the average state of the system defined as

\begin{equation}
	\bar{\x} = \frac{\sum_{j = 1}^N k_j  x_j}{N\kk },
\end{equation}

and $\s$ captures the connectivity of the network defined by 

\begin{equation}
	\s = \omega \kappa,
\end{equation}

where $\kappa$ is the average degree of a neighbor,

\begin{equation}
	\kappa = \frac{\sum_{j = 1}^N k_j^2}{N\kk } = \frac{\kkk}{\kk}.
	\label{kappa}
\end{equation}

To capture the steady state of the system we demand relaxation, i.e.\ $\dif \bar{\x} / \dif t = 0$, to obtain

\begin{equation}
	0 = M_0(\bar{\x}) + \s M_1(\bar{\x})M_2(\bar{\x}).
	\label{MFfp}
\end{equation}

We consider dynamics for which $\bar{\x}=0$ is a steady state of the system and thus a solution of the last equation. This stable solution represents an \emph{inactive} stable state of the system. The additional solution is obtained by the relation,

\begin{equation}
	\s = \frac{-M_0(\bar{\x})}{M_1(\bar{\x})M_2(\bar{\x})},
	\label{sXfree}
\end{equation}

providing the system state for a given connectivity $\s$. This nonzero solution represents an \emph{active} stable state of the system. Together with the zero solution, a phase transition appears.

%

For instance, in the SIS model mentioned above the equation for the steady states, Eq.\ \eqref{SIS}, is

\begin{equation}
	0 = -\alpha \bar{\x} + \s (1-\bar{\x})\bar{\x}.
	\label{SISfp}
\end{equation}

\begin{figure}[h]
	\centering
	\includegraphics[width=0.35\linewidth]{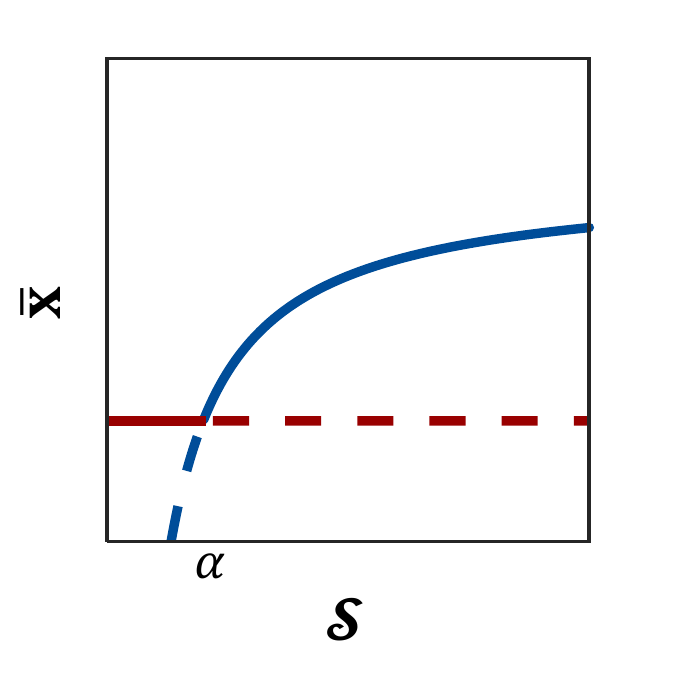}
	\caption{{\bf Epidemic dynamics.} For SIS model for epidemic spreading, Eq.~\eqref{SIS}, the mean-field equation for the fixed points, Eq.\ \eqref{SISfp}, has two solutions; inactive (red, $\bar{\x}=0$) and active (blue, $\bar{\x}>0$). The critical point is $\s_c=\alpha$ at which a continuous phase transition occurs. The dashed lines represent unstable fixed points while the continuous lines represent stable states of the system.} 
	\label{FigSIS}
\end{figure}

One can see that $\bar{\x}=0$ is a solution, and the nonzero solution is $\bar{\x}=1-\alpha/\s$. Therefore, a continuous phase transition is obtained at the intersection between the two solutions, which occurs at $\s_c=\alpha$, as seen in Fig.\ \ref{FigSIS}.

\clearpage

\section{Controlling network dynamics}

To control network dynamics we force a set of nodes, $\F$, (a fraction $\rho$ of the system) to have a constant high activity value $\Delta$, while all the rest in the complementary set, $\D$, are governed by the original dynamics, 

\begin{equation}
	\left\{
	\begin{array}{cclr}
		x_i &=& \Delta & i \in \mathcal{F}
		\\[7pt]
		\dod{x_i}{t} &=& M_0(x_i) + \omega \displaystyle \sum_{j = 1}^N A_{ij} M_1(x_i) M_2(x_j) \hspace{5mm} & i \in \D 
	\end{array}
	\right. .
	\label{ForcedDynamics}
\end{equation} 

In this paper, we assume that the set of controlled nodes, $\F$, is selected randomly.
We observe the state of the unforced nodes, i.e.\ the set $\D$.  

We present here briefly the analysis developed in Ref.\ \cite{sanhedrai2022sustaining} to find the steady state of a controlled system.

We assume that all dynamic free nodes, in $\D$, have approximately a similar neighborhood, namely a similar distribution of activities values $x_j$. This allows us to use the mean-field approach. For a controlled system, Eq.\ \eqref{Dynamics} becomes, for the dynamic free nodes, 

\begin{equation}
	\dod{x_i}{t} = M_0(x_i) + \omega M_1(x_i) \bigg(\sum_{j \in \D} A_{ij}  M_2(x_j) + \sum_{j \in \F} A_{ij}  M_2(\Delta) \bigg) .
	\label{MF1forced}
\end{equation}

We denote $k_{i}^{\D\to\D}$ as the number of $i$'s free neighbors, i.e.\ within $\D$, and $k_{i}^{\D\to\F}$ as the number of $i$'s forced neighbors, i.e.\ in $\F$, where $i\in\D$ is a dynamic free node. A similar mean field approximation as above, Eqs.\ \eqref{MF1}-\eqref{MF4}, yields

\begin{equation}
	\dod{x_i}{t} = M_0(x_i) + \omega M_1(x_i) \bigg(k_i^{\D\to\D} M_2(\bar{\x}) + k_i^{\D\to\F}  M_2(\Delta) \bigg) ,
	\label{MF2forced}
\end{equation}

where $\bar{\x}$ is the average state over free neighbors of free nodes, defined by

\begin{equation}
	\bar{\x} = \frac{\sum_{j \in \D} k_j^{\D\to\D} x_j}{\sum_{j \in \D} k_j^{\D\to\D} }.
	\label{XDef}
\end{equation}

Performing this averaging on both sides of Eq.\ \eqref{MF2forced}, and assuming again the mean-field approximation to insert the averaging into the functions, and to separate the average into the factors of a product, we get

\begin{equation}
	\dod{\bar{\x}}{t} = M_0(\bar{\x}) + \omega M_1(\bar{\x}) \Big( \kappa_{\D\to\D} M_2(\bar{\x}) + \kappa_{\D\to\F}  M_2(\Delta) \Big) ,
	\label{dxdtMF}
\end{equation}

where $\kappa_{\D\to\D}$ is the average degree within $\D$ of a node following a link within $\D$,

\begin{equation}
	\kappa_{\D\to\D} = \frac{\sum_{j \in \D} k_j^{\D\to\D} k_j^{\D\to\D} }{\sum_{j \in \D} k_j^{\D\to\D} },
\end{equation}

and $\kappa_{\D\to\F}$ is the average degree within $\F$ of a node following a link within $\D$,

\begin{equation}
	\kappa_{\D\to\F} = \frac{\sum_{j \in \D} k_j^{\D\to\D} k_j^{\D\to\F} }{\sum_{j \in \D} k_j^{\D\to\D} }.
\end{equation}

We find, next, the last two quantities for our case of random selection of controlled nodes. The average degree of a node following link in $\D$ is just the mean degree of a node following an arbitrary link over the network, which is $\kappa$, Eq.\ \eqref{kappa}. Since there is no difference between the degrees of forced and unforced nodes, the probability of any node to belong to $\F$ is just $\rho$, the fraction of the forced nodes. When node is approached by a link within $\D$, it certainly has one neighbor in $\D$. The rest $\kappa-1$ neighbors, split into $(\kappa-1)\rho$ forced nodes and $(\kappa-1)(1-\rho)$ free nodes in average. Therefore,

\begin{equation}
	\begin{aligned}
		\kappa_{\D\to\D} &= 1+(\kappa-1)(1-\rho),
		\\[5pt]
		\kappa_{\D\to\F} &= (\kappa-1)\rho.
	\end{aligned}
	\label{kappaDDkappaDF}
\end{equation}

When $\kappa$ is large and $\rho$ is small, Eq.\ \eqref{dxdtMF} becomes by substituting Eq.\ \eqref{kappaDDkappaDF} to

\begin{equation}
	\dod{\bar{\x}}{t} = M_0(\bar{\x}) + \s M_1(\bar{\x}) \Big( (1-\rho) M_2(\bar{\x}) + \rho M_2(\Delta) \Big),
	\label{dxdtforced}
\end{equation}

which is Eq.\ (15) in the main text. $\s$ is the connectivity defined as 
\begin{equation}
	\s=\omega\kappa. 
\end{equation}
For the steady state we demand a vanishing of the derivative to obtain Eq.\ (3) in the main text,

\begin{equation}
	\s = \frac{-M_0(\bar{\x})}{M_1(\bar{\x}) \big( (1-\rho) M_2(\bar{\x}) + \rho M_2(\Delta) \big)}.
	\label{sXforced}
\end{equation}

Note that when $\rho=0$ this equation recovers Eq.\ \eqref{sXfree} of a free system.

\clearpage

\section{Classification of the transition}

After obtaining in the previous Section the general formulas for the system states, we find in this Section how the leading orders of the dynamical functions affect the nature of the transition. This is also an introduction to the next Sections where we evaluate the critical exponents from the leading orders of the dynamic functions.

First, we analyze the phase transition of a free system without external control.
We assume there exist the following Hahn power series expansions \cite{Hahn1995} for the dynamical functions in Eq.\ \eqref{Dynamics},

\begin{equation}
	\begin{aligned}
		M_0(x) &= \sum\limits_{n=0}^{\infty}a_nx^{\Gamma _n} ,
		\\
		Q(x) = M_1(x)M_2(x) &= \sum\limits_{n=0}^{\infty}b_nx^{\Pi _n} ,
		\\
		M_1(x) &= \sum\limits_{n=0}^{\infty}c_nx^{\Lambda _n}	,
		\label{series}
	\end{aligned}
\end{equation}

where $\Gamma _n$, $\Pi _n$ and $\Lambda _n$ are increasing non-negative series, and $a_0,b_0,c_0\ne0$. The exponents can be fractional in contrast to Taylor series.

Next, we find several characteristics of these series required for a continuous phase transition between an inactive ($\bar{\x}=0$) stable state for a weak connectivity $\s<\s_c$ and an active ($\bar{\x}>0$) stable state for a strong connectivity $\s>\s_c$.

These characteristics as explained below are

\begin{equation}
	\begin{array}{c c c c l}
		\Gamma_0 &=&\Pi_0 &>&0,
		\\[5pt]
		0 &\le& \Lambda_0 &<& \Pi_0,
		\\[5pt]
		a_0 &<& 0 &<& b_0,
		\\[5pt]
		& & 0 &<& c_0M_2(\Delta).
	\end{array}
\end{equation}

Since we consider dynamics for which $\bar{\x}=0$ is a stable state, therefore $\bar{\x}=0$ fulfills Eq.\ \eqref{MFfp} for the range $0\le\s\le\s_c$ including $\s=0$. Thus, we conclude
\begin{equation}
	\begin{aligned}
		M_0(0) &= 0,
		\\[5pt]
		M_1(0)M_2(0) &= 0.
	\end{aligned}
	\label{M00M10}
\end{equation}
This implies that $\Gamma_0>0$ and $\Pi_0>0$. 

Since $\bar{\x}=0$ is a stable state for $0\le\s<\s_c$, we demand that $\dif \bar{\x}/ \dif t < 0$ for $\s=0$ and $\bar{\x}\to0^+$, that is,
\begin{equation}
	a_0\bar{\x}^{\Gamma_0} + ... < 0,
\end{equation}
yielding $a_0<0$.

To capture the transition point, we search where the active stable solution of Eq.\ \eqref{sXfree} meets the inactive solution $\bar{\x}=0$.
To this end, we expand Eq.\ \eqref{sXfree} in the limit of $\bar{\x}\to0^+$.

\begin{equation}
	\s = \frac{ -M_0(\bar{\x})}{M_1(\bar{\x})M_2(\bar{\x})} = \frac{ -M_0(\bar{\x})}{Q(\bar{\x})} = \frac{-a_0{\bar{\x}}^{\Gamma_0} +...}{b_0{\bar{\x}}^{\Pi_0}+...} \sim \frac{-a_0}{b_0}{\bar{\x}}^{\Gamma_0-\Pi_0}.
	\label{sXfreeScaling}
\end{equation}

Here we distinguish between three possible cases:
\begin{enumerate}
\item $\Gamma_0-\Pi_0>0$. In this case, $\bar{\x}\to0$ yields $\s\to0$, which indicates there is no transition. Thus, we will not discuss this case.
\item $\Gamma_0-\Pi_0<0$. In this case, $\bar{\x}\to0$ yields $\s\to\infty$, which indicates there is no \emph{continuous} transition. However, it includes cases where there is an abrupt transition. Ref.\ \cite{sanhedrai2022sustaining} explores this case in detail. 
\item $\Gamma_0-\Pi_0=0$. In this case, $\bar{\x}\to0$ yields $\s\to\s_c>0$, which indicates a continuous phase transition which is our focus in our paper. 
\end{enumerate}
Therefore, we assume in our paper that $\Gamma_0=\Pi_0$. Note that the critical point obtained from Eq.\ \eqref{sXfreeScaling} is at $\s_c=-a_0/b_0$, and for $\s_c$ to be positive, we demand $b_0>0$.

As an example, let us observe the Michaelis-Menten model for regulatory dynamics \cite{Karlebach2008}, captured by

\begin{equation}
	\dod{x_i}{t} = -Bx_i^a + \omega \sum_{j=1}^{N} \m Aij \frac {x_j^h}{1+x_j^h}.
	\label{MM}
\end{equation}

\begin{figure}[h]
	\centering
	\includegraphics[width=0.9\linewidth]{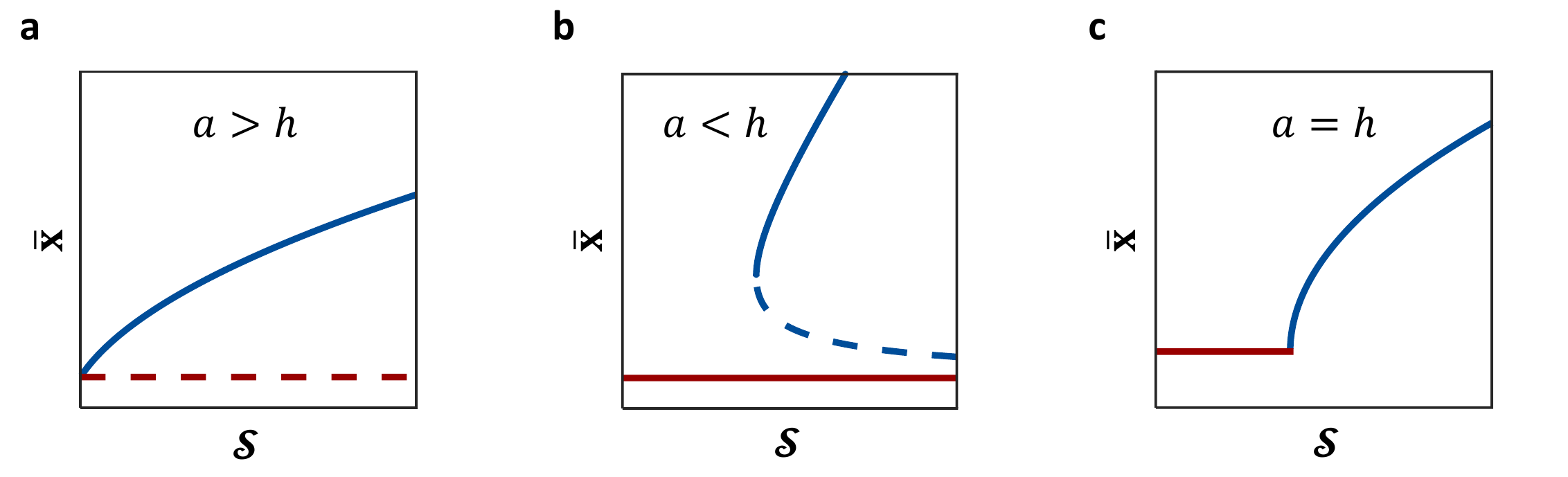}
	\caption{{\bf Regulatory dynamics} (a) For $a>h$, the active solution (blue, $\bar{\x}>0$) approaches zero at $\s=0$, thus there is no phase transition. The dashed red line represents an unstable inactive state. (b) If $a<h$ the non-zero solution (blue, $\bar{\x}>0$) approaches zero at $\s\to\infty$, which implies an abrupt phase transition between the inactive (red) and the active (blue) states. The dashed line represents an unstable fixed point. (c) When $a=h$ the active (blue, $\bar{\x}>0$) stable state approaches zero at the critical point $\s_c>0$. Therefore, a continuous phase transition is obtained. This is the case we analyze in this paper.}
	\label{Fig3ah}
\end{figure}

In this model, $\Gamma_0=a$ and $\Pi_0=h$. See Fig.\ \ref{Fig3ah} for solution of Eq.\ \eqref{sXfree} in the three mentioned above cases which are expressed here in the relation between $a$ and $h$. One can see that only the case of $a=h$ provides a continuous phase transition, which is our focus, thus we set in this paper $a=h$.

Next, we show that an addition of control increases the stable states and smooths the transition, such that $\bar{\x}=0$ is no more a stable state of the system. As such, controlling system dynamics acts as an external field in a continuous phase transition. For this purpose, we expand Eq.\ \eqref{sXforced} in the limit of $\bar{\x}\to0$ where $\rho>0$, and recalling $\Pi_0=\Gamma_0$, we obtain

\begin{equation}
	\s=\frac{ -M_0(\bar{\x})}{(1-\rho)M_1(\bar{\x})M_2(\bar{\x})+\rho M_1(\bar{\x})M_2(\Delta)} \sim \frac{-a_0{\bar{\x}}^{\Gamma_0}}{(1-\rho)b_0{\bar{\x}}^{\Gamma_0}+\rho c_0{\bar{\x}}^{\Lambda_0}M_2(\Delta)}  .
	\label{}
\end{equation}

If $\Lambda_0>\Pi_0$, then the limit of $\bar{\x}\to0$ implies that a transition still exists, it just moves slightly to
$\s_c=-a_0/b_0/(1-\rho)$.
If $\Lambda_0=\Pi_0$, again there still exists a shifted critical point $\s_c=-a_0/((1-\rho)b_0+\rho c_0M_2(\Delta))$. Since we wish to explore a control that acts as an external field that smooths the transition and eliminates the critical point, we assume in our paper $\Lambda_0<\Pi_0$ that yields

\begin{equation}
	\s \sim
		\frac{-a_0}{\rho c_0 M_2(\Delta)}{\bar{\x}}^{\Gamma_0-\Lambda_0},
	\label{}
\end{equation}

where the exponent is positive, and hence $\bar{\x}\to0^+$ leads to $\s\to0^+$, indicating that the control eliminates the phase transition and smooths the function analogously to the effect of an external field. Since $\s$ and $\bar{\x}$ are positive, we conclude that $c_0M_2(\Delta)>0$.

\clearpage

\section{Critical exponents}

As shown in the previous Section, we consider dynamical processes that exhibit a continuous phase transition, and controlling the dynamics of a fraction of nodes acts as an external field. Therefore, we define here $\bar{\x}$ as the order parameter, $\s$ as the tuning parameter, and $\rho$ as the external field. From these definitions, we derive the corresponding critical exponents related to this phase transition.

\subsection{The exponent $\beta$}

First, we find the critical exponent $\beta$, defined by

\begin{equation}
	\bar{\x} \sim ( \s-\s_c ) ^{\beta},
	\label{betaDef}
\end{equation}

where $\s\to\s_c$.
The critical point, $\s_c$ is obtained from Eq.\ \eqref{sXfreeScaling} by recalling that $\Gamma_0=\Pi_0$ and taking $\bar{\x}\to0$. This provides $\s_c=-a_0/b_0$.
To obtain the relation between $\bar{\x}$ and $\s-\s_c$, we use Eq.\ \eqref{sXfree} and \eqref{series} where $\Gamma_0=\Pi_0$,

\begin{equation}
	\begin{split}
		\s-\s_c &= \frac{-a_0{\bar{\x}}^{\Gamma _0}-a_1{\bar{\x}}^{\Gamma_1}+...}{b_0{\bar{\x}}^{\Gamma_0}+b_1{\bar{\x}}^{\Pi_1}+...} + \frac{a_0}{b_0} = \frac{-a_0}{b_0} \bigg( \frac{1+\frac{a_1}{a_0}{\bar{\x}}^{\Gamma_1-\Gamma_0}+...}{1 + \frac{b_1}{b_0}{\bar{\x}}^{\Pi_1-\Gamma_0}+...} - 1 \bigg) \\[7pt]
		& = \frac{-a_0}{b_0} \frac{\frac{a_1}{a_0}{\bar{\x}}^{\Gamma_1-\Gamma_0}-\frac{b_1}{b_0}{\bar{\x}}^{\Pi_1-\Gamma_0}+...}{1 + \frac{b_1}{b_0}{\bar{\x}}^{\Pi_1-\Gamma_0}+...} \sim {\bar{\x}}^{\min\{\Gamma_1,\Pi_1\}-\Gamma_0} .
		\label{smscXscaling}
	\end{split}
\end{equation}

Therefore,
\begin{equation}
	\bar{\x} \sim ( \s-\s_c ) ^{\frac{1}{\min\{\Gamma_1,\Pi_1\}-\Gamma_0}},
\end{equation}
implying
\begin{equation}
	\beta = \frac{1}{\min\{\Gamma_1,\Pi_1\}-\Gamma_0}.
\end{equation}
By defining 
\begin{equation}
	m = \min\{\Gamma_1,\Pi_1\},
	\label{m}
\end{equation}
we obtain
\begin{equation}
	\beta = \frac{1}{m-\Gamma_0}.
	\label{beta}
\end{equation}

Regarding the definition of $m=\min\{\Gamma_1,\Pi_1\}$, we need to clarify that if $\Gamma_1$ does not exist then simply $m=\Pi_1$, and similarly if $\Pi_1$ does not exist, then $m=\Gamma_1$. For the particular case in which $\Gamma_1=\Pi_1$ and also $a_1/a_0=b_1/b_0$, the lead terms in Eq.\ \eqref{smscXscaling} cancel each other, and thus $m$ used in Eq.\ \eqref{beta} is larger than $\Gamma_1=\Pi_1$. However, we do not treat this singular case since this is not the situation in either of our examples. 

Another issue we can learn from Eq.\ \eqref{smscXscaling} is that the coefficient of the leading term of the nominator should be positive since $\s-\s_c>0$ for $\bar{\x}>0$. Thus, if $\Gamma_1<\Pi_1$ then $a_1/a_0>0$, but we already know that $a_0<0$, thus we conclude that $a_1<0$. Similarly, if $\Pi_1<\Gamma_1$, then $b_1/b_0<0$, but we already know that $b_0>0$, thus we conclude that $b_1<0$. In case that $\Gamma_1=\Pi_1$ we demand $a_1/a_0-b_1/b_0>0$, and if this is zero, then the next order should be considered. However, we omit this singular case as mentioned above.

\subsection{The exponent $\delta$}

Next, the exponent $\delta$ is defined by the dependence of the order parameter on the external field at criticality,

\begin{equation}
	{\bar{\x}}(\s_c) \sim {\rho}^{1/\delta},
	\label{deltaDef}	
\end{equation}

where $\rho\to0$.
In the presence of control, we use Eq.\ \eqref{sXforced} to conclude that at criticality, $\s=\s_c=-a_0/b_0$,

\begin{equation}
		\frac{-a_0}{b_0} = \frac{ -M_0(\bar{\x})}{(1-\rho)M_1(\bar{\x})M_2(\bar{\x})+\rho M_1(\bar{\x})M_2(\Delta)}.
\end{equation}

Isolating $\rho$ gives

\begin{equation}
	\rho = \frac{ M_0(\bar{\x})b_0/a_0 - M_1(\bar{\x})M_2(\bar{\x}) }{-M_1(\bar{\x})M_2(\bar{\x})+ M_1(\bar{\x})M_2(\Delta)} ,
\end{equation}

and by using Eq.\ \eqref{series},

\begin{equation}
	\begin{aligned}		
			\rho & = \frac{\frac{b_0}{a_0}(a_0{\bar{\x}}^{\Gamma _0}+a_1{\bar{\x}}^{\Gamma_1}+...)-(b_0{\bar{\x}}^{\Gamma_0}+b_1{\bar{\x}}^{\Pi_1}+...)}{(c_0{\bar{\x}}^{\Lambda_0}+...)M_2(\Delta)-(b_0{\bar{\x}}^{\Gamma_0}+...)} 
			\\[10pt]
			& = \frac{\frac{a_1b_0}{a_0}{\bar{\x}}^{\Gamma_1}-b_1{\bar{\x}}^{\Pi_1}+...}{c_0M_2(\Delta){\bar{\x}}^{\Lambda_0}-b_0{\bar{\x}}^{\Gamma_0}+...} .
	\end{aligned}
	\label{rhoScaling}
\end{equation}

Since as mentioned above $\Lambda_0<\Pi_0=\Gamma_0$,

\begin{equation}
	 \rho \sim {\bar{\x}}^{\min\{\Gamma_1,\Pi_1\} - \Lambda_0},
\end{equation}

resulting in

\begin{equation}
	{\bar{\x}} \sim {\rho}^{\frac{1}{\min\{\Gamma_1,\Pi_1\} - \Lambda_0}}.
\end{equation}

Therefore,

\begin{equation}
	\delta = \min\{\Gamma_1,\Pi_1\} - \Lambda_0 = m - \Lambda_0 .
	\label{delta}	
\end{equation}

\subsection{The exponent $\gamma$}

We define the susceptibility as the derivative of $\bar{\x}(\s,\rho)$ with respect to $\rho$ at $\rho=0$,

\begin{equation}
	\chi = \pd{ \bar{\x}}{\rho}\bigg|_{\rho=0},
\end{equation}

and the exponent $\gamma$ through

\begin{equation}
	\chi \sim (\s-\s_c)^{-\gamma} ,
\end{equation}

for $\s\to\s_c$.
Eq.\ \eqref{sXforced} can be written as

\begin{equation}
	(1-\rho)M_1(\bar{\x})M_2(\bar{\x}) + \rho M_1(\bar{\x})M_2(\Delta) = -\frac{1}{\s} M_0(\bar{\x}) ,
\end{equation}
or
\begin{equation}
	(1-\rho)Q(\bar{\x}) + \rho M_1(\bar{\x})M_2(\Delta) = -\frac{1}{\s} M_0(\bar{\x}) .
\end{equation}

Let us derive both sides by $\partial /\partial \rho |_{\rho=0}$ and use the chain rule,

\begin{equation}
	-Q(\bar{\x}) + \chi  Q'(\bar{\x})+ M_1(\bar{\x})M_2(\Delta) = -\frac{1}{\s} \chi M_0'(\bar{\x}).
\end{equation}

Isolating $\chi$ gives

\begin{equation}
	\chi = \frac{-Q(\bar{\x})+M_1(\bar{\x})M_2(\Delta)}{ -Q'(\bar{\x})-\dfrac{1}{\s} M_0'(\bar{\x})}	.
\end{equation}

To find the exponent $\gamma$ we expand this equation close to criticality, i.e.\ $\bar{\x}\to0$. We use Eq.\ \eqref{series} to expand the functions with respect to $\bar{\x}$, and Eq.\ \eqref{smscXscaling} to write $\s\sim -a_0/b_0(1+a_1/a_0\bar{\x}^{\Gamma_1-\Gamma_0}-b_1/b_0\bar{\x}^{\Pi_1-\Gamma_0}+...)$. This yields for the leading order

\begin{equation}
	\chi \sim \frac{-b_0\bar{\x}^{\Pi_0} + c_0 M_2(\Delta)\bar{\x}^{\Lambda_0}}{ -b_0\Gamma_0{\bar{\x}}^{\Gamma_0-1}-b_1\Pi_1{\bar{\x}}^{\Pi_1-1} + \dfrac{b_0}{a_0}\dfrac{a_0\Gamma_0{\bar{\x}}^{\Gamma_0-1}+a_1\Gamma_1{\bar{\x}}^{\Gamma_1-1}}{1+a_1/a_0\bar{\x}^{\Gamma_1-\Gamma_0}-b_1/b_0\bar{\x}^{\Pi_1-\Gamma_0}} }	.
\end{equation}

Arranging terms and neglecting non-lead terms gives

\begin{equation}
	\chi \sim \frac{ c_0 M_2(\Delta) \bar{\x}^{\Lambda_0} } { -b_1\Pi_1{\bar{\x}}^{\Pi_1-1} + \frac{a_1b_0}{a_0}\Gamma_1{\bar{\x}}^{\Gamma_1-1} - \frac{b_0a_1}{a_0}\Gamma_0{\bar{\x}}^{\Gamma_1-1}+b_1\Gamma_0{\bar{\x}}^{\Pi_1-1}}	.
\end{equation}

Unifying terms, we get

\begin{equation}
	\chi \sim \frac{ \frac{c_0}{b_0} M_2(\Delta) \bar{\x}^{\Lambda_0} } { \frac{a_1}{a_0}(\Gamma_1-\Gamma_0){\bar{\x}}^{\Gamma_1-1}-\frac{b_1}{b_0}(\Pi_1-\Pi_0){\bar{\x}}^{\Pi_1-1} }.
\end{equation}

Therefore, in a similar way as for Eq.\ \eqref{smscXscaling}, we obtain for the leading order,

\begin{equation}
	\chi \sim  \bar{\x}^{\Lambda_0-\min\{\Gamma_1,\Pi_1\}+1} .
\end{equation}

Substituting $m=\min\{\Gamma_1,\Pi_1\}$, Eq.\ \eqref{m}, and using $\bar{\x}\sim(\s-\s_c)^{\beta}$, Eqs.\ \eqref{betaDef} and \eqref{beta}, we get

\begin{equation}
	\chi \sim  (\s-\s_c)^{\beta(\Lambda_0-m+1)} = (\s-\s_c)^{\frac{\Lambda_0-m+1}{m-1}}.
\end{equation}

As a result, we obtain

\begin{equation}
	\gamma = \frac{m-\Lambda_0-1}{m-\Gamma_0} .
	\label{gamma}
\end{equation}

One can see from Eqs.\ \eqref{beta}, \eqref{delta} and \eqref{gamma} that Widom's identity, $\beta(\delta-1)=\gamma$, is satisfied.

\clearpage

\section{Critical exponents of the transient time}

In this Section, we analyze the time that takes for the system to reach the steady state, i.e.\ the transient time. We define below three exponents related to the transient. One of them is related to the effect of controlling system dynamics as an external field. 

\subsection{The exponent $\varphi$}

The dynamical mean-field equation for a free system, Eq.\ \eqref{MFDynamics}, is

\begin{equation}
	\dod{\bar{\x}}{t} = M_2(\bar{\x}) + \s M_1(\bar{\x}) M_2(\bar{\x}) .
\end{equation}

We analyze this equation at criticality, $\s=\s_c=-a_0/b_0$, and in the limit of large $t$ when $\bar{\x}\to0$,

\begin{equation}
	\dod{\bar{\x}}{t} \approx (a_0\bar{\x}^{\Gamma_0}  + a_1\bar{\x}^{\Gamma_1}+... ) -\frac{a_0}{b_0} (b_0\bar{\x}^{\Pi_0} + b_1\bar{\x}^{\Pi_1}+...) .
\end{equation}

Recalling $\Gamma_0=\Pi_0$,
 
\begin{equation}
	\dod{\bar{\x}}{t} \sim  a_1 \bar{\x} ^ {\Gamma_1} -a_0b_1/b_0 \bar{\x} ^{\Pi_1} \sim -\bar{\x} ^{\min\{\Gamma_1,\Pi_1\}} .
\end{equation}

As above, Eq.\ \eqref{m}, we denote $m = \min\{\Gamma_1,\Pi_1\}$ to get for $\s=\s_c$,

\begin{equation}
	\dod{\bar{\x}}{t} \sim -\bar{\x}^m.
\end{equation}

The solution for this equation varies for different values of $m$. \\
For $m<1$, we get a finite time for $\bar{\x}$ to reach zero.\\
For $m=1$, we get an exponential convergence,

\begin{equation}
	\bar{\x} \sim  e^{-t/\tau}.
\end{equation}

For $m>1$, which is the case in all our examples, the asymptotic behavior for a long time is a power law,

\begin{equation}
	\bar{\x} \sim  t^{-\frac{1}{m-1}}.
\end{equation}

Therefore, we define for the case of $m>1$ the dynamic critical exponent $\varphi$ as

\begin{equation}
	\bar{\x} \sim  t^{-\varphi},
\end{equation}

resulting in

\begin{equation}
	\varphi = \frac{1}{m-1}.
\end{equation}


\subsection{The exponents $\phi$ and $\theta$}

Next, we move to analyze the dynamic convergence to the stable state not at criticality.
We use the mean field dynamic equation for a controlled system, Eq.\ \eqref{dxdtforced}, 

\begin{equation}
	\dod{\bar{\x}}{t} = M_0(\bar{\x}) + \s M_1(\bar{\x}) \Big( (1-\rho) M_2(\bar{\x}) + \rho M_2(\Delta) \Big).
	\label{dxdtforced2}
\end{equation}

We expand it for a large $t$ when $\bar{\x}$ is close to the stable state. Here we consider a stable state which satisfies $\bar{\x}>0$ because $\rho>0$ or $\s>\s_c$. 
The rhs of Eq.\ \eqref{dxdtforced2} vanishes at the fixed point, and since we consider a stable state the derivative of the rhs with respect to $\bar{\x}$ is negative. In addition we assume that the function of rhs is analytic for $\bar{\x}>0$ and as such it has a Taylor expansion. Thus, we denote $\bar{\x}_{\infty}=\bar{\x}(t\to\infty)$ as the stable state, and $\delta\bar{\x}=\bar{\x}(t)-\bar{\x}_{\infty}$ as the distance from the stable state, to obtain

\begin{equation}
	\dod{\delta\bar{\x}}{t} \sim \left[ M_0'(\bar{\x}_{\infty}) + \s (1-\rho)Q'(\bar{\x}_{\infty}) + \s \rho M_1'(\bar{\x}_{\infty}) M_2(\Delta))\right] \delta\bar{\x} .
	\label{}
\end{equation}

This yields an exponential convergence to the steady state with decay time $\tau(\s,\rho)$, where

\begin{equation}
	\dod{\delta\bar{\x}}{t} \sim -\frac{1}{\tau(\s,\rho)} \delta\bar{\x} ,
	\label{}
\end{equation}

and 

\begin{equation}
	\tau = -\frac{1}{ M_0'(\bar{\x}_{\infty}) + \s (1-\rho) Q'(\bar{\x}_{\infty}) + \s \rho M_1'(\bar{\x}_{\infty}) M_2(\Delta))} .
	\label{tauDef}
\end{equation}

As shown above, at criticality, i.e.\ $\rho=0$ and $\s=\s_c$, the convergence behaves as a power law rather than exponential (for $m>1$), therefore we expect that 

\begin{equation}
	\tau(\s\to\s_c,\rho\to0) \to \infty.
\end{equation}

We show below that this divergence is as power law and we correspondingly define two exponents as follows,

\begin{eqnarray}
	\label{phiDef}	
	\tau(\s\to\s_c^+,\rho=0) & \sim & {(\s-\s_c)}^{-\phi} ,
	\\[8pt]
	\label{thetaDef}
	\tau(\s=\s_c,\rho\to0) & \sim & {\rho}^{-\theta} .
\end{eqnarray}

\subsubsection*{The exponent $\phi$}

Let us start with the first scaling. We substitute in Eq.\ \eqref{tauDef} $\rho=0$ and obtain

\begin{equation}
	\tau = -\frac{1}{M_0'(\bar{\x}_{\infty})+\s Q'(\bar{\x}_{\infty})},
	\label{tauRho0}
\end{equation}

and for $\s\to\s_c=-a_0/b_0$, it holds that $\bar{\x}_{\infty}\to0$ and according to
Eq.\ \eqref{smscXscaling} $\s\sim -a_0/b_0(1+a_1/a_0\bar{\x}_{\infty}^{\Gamma_1-\Gamma_0}-b_1/b_0\bar{\x}_{\infty}^{\Pi_1-\Gamma_0}+...)$, yielding

\begin{equation}
	\begin{split}
		\tau & \sim -\frac{1}{a_0\Gamma_0\bar{\x}_{\infty}^{\Gamma_0-1}+a_1\Gamma_1\bar{\x}_{\infty}^{\Gamma_1-1}-\frac{a_0}{b_0}(1+\frac{a_1}{a_0}\bar{\x}_{\infty}^{\Gamma_1-\Gamma_0}-\frac{b_1}{b_0}\bar{\x}_{\infty}^{\Pi_1-\Gamma_0}) (b_0\Pi_0\bar{\x}_{\infty}^{\Pi_0-1}+b_1\Pi_1\bar{\x}_{\infty}^{\Pi_1-1})}
		\\[5pt]
		& \sim -\frac{1}{a_1\Gamma_1\bar{\x}_{\infty}^{\Gamma_1-1}-a_0\frac{b_1}{b_0}\Pi_1\bar{\x}_{\infty}^{\Pi_1-1} -a_1\Pi_0\bar{\x}_{\infty}^{\Gamma_1-1}+a_0\frac{b_1}{b_0}\Pi_0\bar{\x}_{\infty}^{\Pi_1-1}}
		\\[5pt]
		& \sim \frac{-\frac{1}{a_0}}{\frac{a_1}{a_0}(\Gamma_1-\Gamma_0)\bar{\x}_{\infty}^{\Gamma_1-1}-\frac{b_1}{b_0}(\Pi_1-\Pi_0)\bar{\x}_{\infty}^{\Pi_1-1} }.
	\end{split}	
\end{equation}

Similarly to Eq.\ \eqref{smscXscaling}, denoting $m=\min\{\Gamma_1,\Pi_1\}$, we conclude that

\begin{equation}
	\tau \sim \bar{\x}_{\infty}^{-(m-1)} \sim (\s-\s_c) ^{-\beta(m-1)} \sim  (\s-\s_c) ^{-\frac{m-1}{m-\Gamma_0}}.
\end{equation}

This yields

\begin{equation}
	\phi = \frac{m-1}{m-\Gamma_0}.
\end{equation}

\subsubsection*{The exponent $\theta$}

Next, we substitute in Eq.\ \eqref{tauDef} $\s=\s_c=-a_0/b_0$, 

\begin{equation}
	\frac{1}{\tau} = - M_0'(\bar{\x}_{\infty}) +\frac{a_0}{b_0} \Big[(1-\rho) Q'(\bar{\x}_{\infty}) + \rho M_1'(\bar{\x}_{\infty}) M_2 (\Delta)\Big],
\end{equation}

and we take the limit of $\rho\to0$ resulting in $\bar{\x}_{\infty}\to0$, and according to Eq.\ \eqref{rhoScaling}
$
\rho \sim b_0/(c_0M_2(\Delta)) \big(a_1/a_0\bar{\x}_{\infty}^{\Gamma_1-\Lambda_0}-b_1/b_0\bar{\x}_{\infty}^{\Pi_1-\Lambda_0}\big) .
$
This yields

\begin{equation}
	\begin{aligned}
		\frac{1}{\tau} & \sim  -a_0\Gamma_0\bar{\x}_{\infty}^{\Gamma_0-1} - a_1\Gamma_1\bar{\x}_{\infty}^{\Gamma_1-1} 
		\\[5pt]
		& +\frac{a_0}{b_0}\Big(1-\frac{b_0}{c_0M_2(\Delta)} \Big(\frac{a_1}{a_0}\bar{\x}_{\infty}^{\Gamma_1-\Lambda_0}-\frac{b_1}{b_0}\bar{\x}_{\infty}^{\Pi_1-\Lambda_0}\Big)\Big) \big(b_0\Pi_0\bar{\x}_{\infty}^{\Pi_0-1}+b_1\Pi_1\bar{\x}_{\infty}^{\Pi_1-1}\big) 
		\\[5pt]
		& + \frac{a_0}{b_0} \frac{b_0}{c_0M_2(\Delta)} \Big(\frac{a_1}{a_0}\bar{\x}_{\infty}^{\Gamma_1-\Lambda_0}-\frac{b_1}{b_0}\bar{\x}_{\infty}^{\Pi_1-\Lambda_0}\Big) c_0 \Lambda_0\bar{\x}_{\infty}^{\Lambda_0-1} M_2(\Delta) .
	\end{aligned}
\end{equation}

Assembling terms gives

\begin{equation}
	\begin{aligned}
		\frac{1}{\tau} 
		& \sim -a_0 \Big( \frac{a_1}{a_0}\big(\Gamma_1-\Lambda_0\big) \bar{\x}_{\infty}^{\Gamma_1-1} - \frac{b_1}{b_0}\big(\Pi_1-\Lambda_0\big)\bar{\x}_{\infty}^{\Pi_1-1} \Big)		
		\\[5pt]
		& + O(\bar{\x}_{\infty}^{\Gamma_1-1+(\Pi_0-\Lambda_0)}) +  O(\bar{\x}_{\infty}^{\Pi_1-1+(\Pi_0-\Lambda_0)}) .
	\end{aligned}
\end{equation}

Recalling $\Lambda_0<\Pi_0<\Pi_1,\Gamma_1$, and using Eqs.\ \eqref{deltaDef} and \eqref{delta} for the exponent $\delta$,
we obtain in a similar way as for Eq.\ \eqref{smscXscaling} the scaling

\begin{equation}
	\tau \sim \bar{\x}_{\infty}^{-(m-1)} \sim  {\rho^{-\frac{m-1}{\delta}}} \sim  {\rho^{-\frac{m-1}{m - \Lambda_0}}} .
\end{equation}

Therefore,

\begin{equation}
	\theta = \frac{m-1}{m-\Lambda_0}.
\end{equation}

\clearpage

\section{Dynamical models}

In this section, we introduce three dynamical models which fall under our framework, and we demonstrate our general results for each one of them.

\subsection{Epidemic dynamics}

We consider SIS model for epidemic spreading, \cite{Barthelemy2005,Hufnagel2004,Dodds2005} where a susceptible node can get infected by an infectious node, and an infectious node might recover and become susceptible again. According to this model, the probability of each agent $i$ to be infectious, $x_i$, evolves in time as

\begin{equation}
	\dod{x_i}{t} = - \alpha x_i + \omega \sum_{j = 1}^N \m Aij (1-x_i) x_j,
	\label{SIS2}
\end{equation}

where $\alpha$ is the recovery rate, and $\omega$ is the infection rate. The product $(1-x_i)x_j$ represents the probability that agent $i$ is susceptible and node $j$ is infectious.
This dynamical model is mapped to the general form of Eq.\ \eqref{Dynamics} by $M_0(x)=-\alpha x$, $M_1(x)=1-x$ and $M_2(x)=x$. The infection rate $\omega$ is the interaction weight in this dynamics. We set $\alpha=1$ for simplicity.
The obtained dynamic mean-field equation, Eq.\ \eqref{MFDynamics}, is

\begin{equation}
	\dod{\bar{\x}}{t} = -\bar{\x} + \s (1-\bar{\x})\bar{\x},
\end{equation}

and for the fixed points,

\begin{equation}
	0 = -\bar{\x} + \s (1-\bar{\x})\bar{\x}.
\end{equation}

The solution $\bar{\x}=0$ holds for any $\s$. The active solution is $\bar{\x}(\s)=1-1/\s$. The two states meet at $\s_c=1$, indicating a continuous phase transition.

The series of the relevant functions in Eq.\ \eqref{series} are

\begin{equation}
	\begin{aligned}
		M_0(x) & = -x ,
		\\[5pt]
		M_1(x)M_2(x) & = x - x^2 ,
		\\[5pt]
		M_1(x) & = 1 - x .
		\label{}
	\end{aligned}
\end{equation}

Therefore, $\Gamma_0=1$ and $\Gamma_1$ does not exist, $\Pi_0=1$ and $\Pi_1=2$, and $\Lambda_0=0$. Hence, $m=\min\{\Gamma_1,\Pi_1\}=2$, because $\Gamma_1$ does not exist. Thus, the equilibrium critical exponents are

\begin{equation}
	\begin{array}{c c c c c}
		\beta & = & \dfrac{1}{m-\Gamma_0}  & = & 1,
		\\[10pt]
		\delta & = & m-\Lambda_0 & = & 2,
		\\[10pt]
		\gamma & = & \dfrac{m-\Lambda_0-1}{m-\Gamma_0} & = & 1.
	\end{array}
\end{equation}

The transient critical exponents are

\begin{equation}
	\begin{array}{c c c c c}
		\varphi & = & \dfrac{1}{m-1} & = & 1 ,
		\\[12pt]
		\phi & = & \dfrac{m-1}{m-\Gamma_0} & = &  1,
		\\[12pt]
		\theta & = & \dfrac{m-1}{m-\Lambda_0} & = & \dfrac{1}{2}.
	\end{array}
\end{equation}

\subsection{Regulatory}

Here we study regulatory dynamics according to Michaelis-Menten model \cite{Karlebach2008}, captured by

\begin{equation}
	\dod{x_i}{t} = -Bx_i^a + \omega \sum_{j=1}^{N} A_{ij} \frac {x_j^h}{1+x_j^h}.
\end{equation}

Under this framework $M_0(x_i) = -B x_i^a$, describing degradation ($a = 1$), dimerization ($a = 2$) or a more complex bio-chemical depletion process (fractional $a$), occurring at a rate $B$; without loss of generality we set here $B = 1$. The activation interaction is captured by the Hill function of the form $M_1(x_i) = 1$, $M_2(x_j) = x_j^h/(1+x_j^h)$, a \textit{switch-like} function that saturates to $M_2(x_j) \rightarrow 1$ for large $x_j$, representing node $j$'s positive, albeit bounded, contribution to node $i$ activity, $x_i(t)$. 

We set $h=a$ to get a continuous transition as shown above in Fig.\ \ref{Fig3ah}. The obtained dynamic mean-field equation, Eq.\ \eqref{MFDynamics}, is

\begin{equation}
	\dod{\bar{\x}}{t} = -\bar{\x}^a + \s \frac {\bar{\x}^a}{1+\bar{\x}^a},
\end{equation}

and for the fixed points,

\begin{equation}
	0 = -\bar{\x}^a + \s \frac {\bar{\x}^a}{1+\bar{\x}^a}.
\end{equation}

The solution $\bar{\x}=0$ holds for any $\s$. The active solution is $\bar{\x}(\s)=(\s-1)^{1/a}$. The two states meet at $\s_c=1$, indicating a continuous phase transition.

The series of the relevant functions in Eq.\ \eqref{series} are

\begin{equation}
	\begin{aligned}
		M_0(x) & = -x^a ,
		\\[5pt]
		M_1(x)M_2(x) & = x^a - x^{2a} + ... ,
		\\[5pt]
		M_1(x) & = 1 .
		\label{}
	\end{aligned}
\end{equation}

Therefore, $\Gamma_0=a$ and $\Gamma_1$ does not exist, $\Pi_0=a$ and $\Pi_1=2a$, and $\Lambda_0=0$. Hence, $m=\min\{\Gamma_1,\Pi_1\}=2a$, because $\Gamma_1$ does not exist. Thus, the equilibrium critical exponents are

\begin{equation}
	\begin{array}{c c c c c}
		\beta & = & \dfrac{1}{m-\Gamma_0} & = & \dfrac{1}{a},
		\\[12pt]
		\delta & = & m-\Lambda_0 & = & 2a,
		\\[8pt]
		\gamma & = & \dfrac{m-\Lambda_0-1}{m-\Gamma_0} & = & 2-\dfrac{1}{a}.
	\end{array}
\end{equation}

The transient critical exponents are

\begin{equation}
	\begin{array}{c c c c c}
		\varphi & =& \dfrac{1}{m-1} &=& \dfrac{1}{2a-1} ,
		\\[12pt]
		\phi & =& \dfrac{m-1}{m-\Gamma_0} &=& 2-\dfrac{1}{a},
		\\[12pt]
		\theta & =& \dfrac{m-1}{m-\Lambda_0} &=& 1-\dfrac{1}{2a}.
	\end{array}
\end{equation}

\subsection{Opinion dynamics}

Our final example is a model for opinion dynamics \cite{baumann2020modeling},
\begin{equation}
	\dod{x_i}{t} = -x_i + \omega\sum\limits_{j=1}^{N}A_{ij} \tanh (\alpha x_j).
	\label{opinions}
\end{equation}
The sign of $x_i$ describes the agent $i$’s qualitative stance towards a binary issue of choice (\textit{e.g.}\ the preference between two candidates). The absolute value of $x_i$ quantifies the strength of this opinion, or the convincing level.
This model treats opinion dynamics as a
purely collective, self-organized process without any intrinsic individual preferences. Hence, the opinions of agents
lacking social interactions decay toward the neutral state 0, which is ruled by the self-dynamics function, $M_0(x_i)=-x_i$.
The interaction $ij$ is captured by $M_1(x_i)=1$ and $M_2(x_j)= \tanh(\alpha x_j)$. 
This odd nonlinear shape guarantees that an agent $j$ influences others in the direction of its own opinion’s sign,  with a level that increases monotonically with its convincing level,
albeit with saturation since the social influence of extreme opinions is capped.
Thus, $\Gamma_0=1$ and $\Gamma_1$ does not exist, $\Pi_0=1$, and $\Pi_1=3$, and $\Lambda_0=0$. We set $\alpha=1$ for simplicity.

 The obtained dynamic mean-field equation, Eq.\ \eqref{MFDynamics}, is

\begin{equation}
	\dod{\bar{\x}}{t} = -\bar{\x}+ \s \tanh (\bar{\x}),
\end{equation}

and for the fixed points,

\begin{equation}
	0 = -\bar{\x}+ \s \tanh (\bar{\x}).
	\label{opinionsFp}
\end{equation}

\begin{figure}[h]
	\centering
	\includegraphics[width=0.35\linewidth]{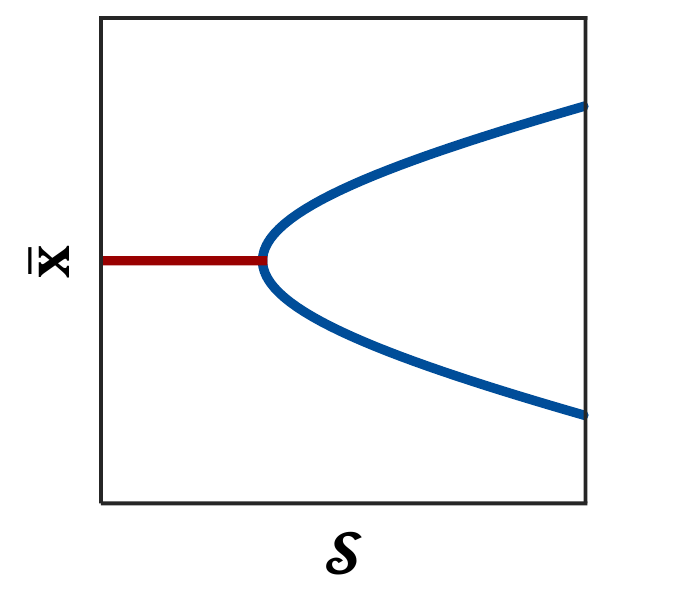}
	\caption{{\bf Opinion dynamics.} For opinion dynamics, \eqref{opinions}, the mean-field equation for the fixed points, Eq.\ \eqref{opinionsFp}, has two solutions; inactive (red, $\bar{\x}=0$) and active (blue, $\bar{\x}\ne0$). The critical point is $\s_c=1$ at which a continuous phase transition occurs. In this paper we consider the positive branch and analyze how a positive control ($\Delta>0$) impacts it.} 
	\label{FigOpinion}
\end{figure}

The solution $\bar{\x}=0$ holds for any $\s$. The active solution satisfies $\s(\bar{\x})=\bar{\x}/\tanh(\bar{\x})$. The two states meet at $\s_c=1$, indicating a continuous phase transition as shown in Fig.\ \ref{FigOpinion}.

The series of the relevant functions in Eq.\ \eqref{series} are

\begin{equation}
	\begin{aligned}
		M_0(x) & = -x ,
		\\[5pt]
		M_1(x)M_2(x) & = x - \frac{x^{3}}{3} + ... ,
		\\[5pt]
		M_1(x) & = 1	.
		\label{}
	\end{aligned}
\end{equation}

Therefore, $\Gamma_0=1$ and $\Gamma_1$ does not exist, $\Pi_0=1$ and $\Pi_1=3$, and $\Lambda_0=0$. Hence, $m=\min\{\Gamma_1,\Pi_1\}=\Pi_1=3$, because $\Gamma_1$ does not exist. Thus, the equilibrium critical exponents are

\begin{equation}
	\begin{array}{c c c c c}
		\beta & = & \dfrac{1}{m-\Gamma_0} & = & \dfrac{1}{2},
		\\[14pt]
		\delta & = & m-\Lambda_0 & = & 3,
		\\[8pt]
		\gamma & = & \dfrac{m-\Lambda_0-1}{m-\Gamma_0} & = & 1.
	\end{array}
\end{equation}

These are the same critical exponents as that of Ising model for ferromagnetism in mean-field. This is not surprising since a very similar model is used to model spin dynamics \cite{krapivsky2010kinetic}. 

The transient critical exponents are

\begin{equation}
	\begin{array}{c c c c c}
		\varphi & = & \dfrac{1}{m-1} &=& \dfrac{1}{2} ,
		\\[12pt]
		\phi & =& \dfrac{m-1}{m-\Gamma_0} &=& 1,
		\\[12pt]
		\theta & =& \dfrac{m-1}{m-\Lambda_0} &=& \dfrac{2}{3}.
	\end{array}
\end{equation}

\clearpage

\bibliographystyle{unsrt}
\bibliography{bibliography}